\documentclass[jmp,showpacs,showkeys,preprint,nofootinbib]{revtex4}
\def\be{\begin{equation}}
\def\ee{\end{equation}}
\def\beq{\begin{eqnarray}}
\def\eeq{\end{eqnarray}}
\def\sch{Schr\"odinger}
\def\se{\sch{} equation}

\usepackage[makeroom]{cancel}
\usepackage{comment}
\usepackage{amsmath}
\usepackage{amssymb}
\usepackage{float}
\usepackage{xcolor}
\usepackage{color,soul}

\usepackage{hyperref}
\usepackage{todonotes}

\usepackage{graphicx}
\usepackage{color}
\usepackage{subcaption}

\begin{document}		
	\title{Analogue black string in a quantum harmonic oscillator}
    
\author{Matheus E. Pereira$^{1,2}$, and Alexandre G. M. Schmidt$^{1,2}$}
\affiliation{$^1$Instituto de Ci\^encias Exatas, Universidade Federal Fluminense,\\ 
	27213-145 Volta Redonda --- RJ, Brazil}
\affiliation{$^2$Programa de P\'os Gradua\c{c}\~ao em F\'isica, Instituto de F\'\i sica, Universidade Federal Fluminense,\\
	24210-346 Niter\'oi --- RJ, Brazil}

\begin{abstract}%

For a scalar particle without self-interaction or backreaction from the space-time background, the dynamics are governed by the Klein-Gordon equation. In this work, we write the exact solution of this equation in the background of a chargeless, static black string in terms of the biconfluent Heun function. In this curious system, we are able to explore what happens if we have negative values for the masses. The eigenvalue problem provides complex energy values for the particle, which may indicate the presence of quasinormal modes. We show a simple quantum system that can imitate the particle in the black string background, whose solutions are also applications of the biconfluent Heun function.

\end{abstract}
	
	
\maketitle
	
\section{Introduction}
The most general asymptotically flat and stationary solution of the Einstein–Maxwell equations is the Kerr--Newman solution, which, according to Hawking's topology theorem, also manifests spherical symmetry at the horizon \cite{hawking, thermo-kerr}. Solutions of Einstein's equations whose event horizon surface does not possess spherical symmetry are thus called \cite{griff} topological black holes, like the black membrane, black string, and toroidal black holes.  

Topological black holes are often found in higher--dimensional or non-asymptotically flat spacetimes, like an AdS spacetime \cite{cisterna-ads}, and are unstable \cite{kudoh} under scalar perturbations, which does not make them any less interesting from the theoretical point of view. Indeed, there have been reports on rotating black strings \cite{lemos-black-string, saifullah}, hairy black strings \cite{herdeiro}, and even a type of stable black string \cite{cisterna-ads}. 

Unfortunately, there seems to exist little progress in the analogue gravity programme \cite{carla} related to the black string, except for a connection in Obukhov's work relating a property of classical confinement of cylindrically symmetric solutions for SU($2$) Yang--Mills gauge theory and an analogue property of black strings \cite{lemos94, lemos96}. The importance of constructing analogue models of gravity has been thoroughly reviewed by \cite{carla}, and the mission of the present work is to start filling this gap in the literature.

Zel'dovich proposed the first analogue model for a gravitational system. It was an experiment \cite{zeldovich} involving wave scattering by a cylindrical object that would act similarly to a black hole. Zel'dovich also predicted a new effect using this analogue system. Later on, Unruh \cite{unruh} investigating acoustic waves found equations equivalent to those of a scalar field in a pseudo-Riemannian space-time. Recently, there have been investigations into quantum analogue models for the Kerr--de Sitter \cite{esferoide} and the Kerr--Sen \cite{kerr-sen} black holes, using a particle on a spheroid and on a sphere immersed in an electromagnetic field, respectively. Analogue models allow us to visualize hidden features, to investigate inaccessible domains of physical quantities (e.g., negative masses, magnetic monopoles \cite{monopolo-pdm, spin-ice} and dyons \cite{dyon-pdm}) and even unphysical quantities \cite{casanova} (e.g., charge conjugation and time reversal) of target systems.

To add to the analogue gravity programme, in this work we present the exact solutions of the wave equation in the background of a non-rotating, chargeless black string in a static universe in terms (or as an application) of  the biconfluent Heun functions. The black string solution allows both the test particle and the black hole masses to be negative, and we present plots for the normalized probability density of the wavefunction in these cases. Further, we propose a damped quantum harmonic oscillator that can mimic the behavior of the test particle, and we solve the Schr\"{o}dinger equation also using the biconfluent Heun function. 

The structure of this work is divided as follows: in section \ref{secao2}, we dive in search of the exact solutions for the Klein--Gordon equation in the background of our black string. Next, section \ref{secao3} shows our suggestion for an analog black string, using a damped quantum harmonic oscillator. The problem is formulated using a one-dimensional system of coupled springs. A table is shown, where we present the mapping of the two systems into each other. We conclude in section \ref{secao4} with our final remarks and future perspectives.
\section{Scalar particle in the static gravitational background of a chargeless black string}\label{secao2}
\begin{figure}[H]
    \begin{subfigure}{.5\textwidth}
  \centering
  \includegraphics[width=0.9\linewidth]{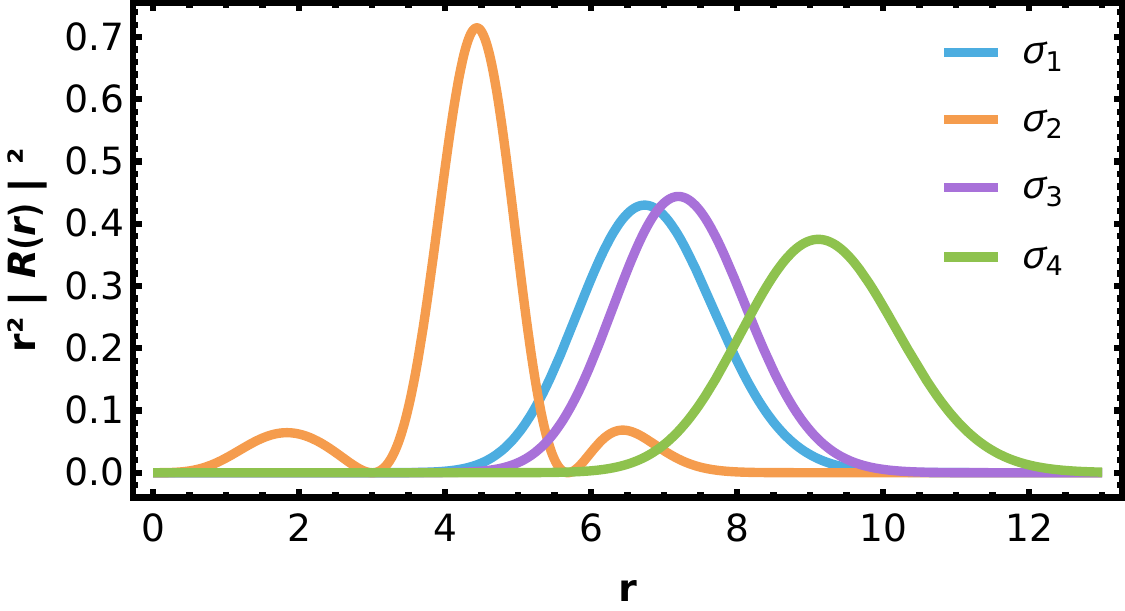}
  \caption{$M = -1$, $\mu_0 = 1$.}
  \label{fig:sfig-a}
\end{subfigure}%
\begin{subfigure}{.5\textwidth}
  \centering
  \includegraphics[width=0.9\linewidth]{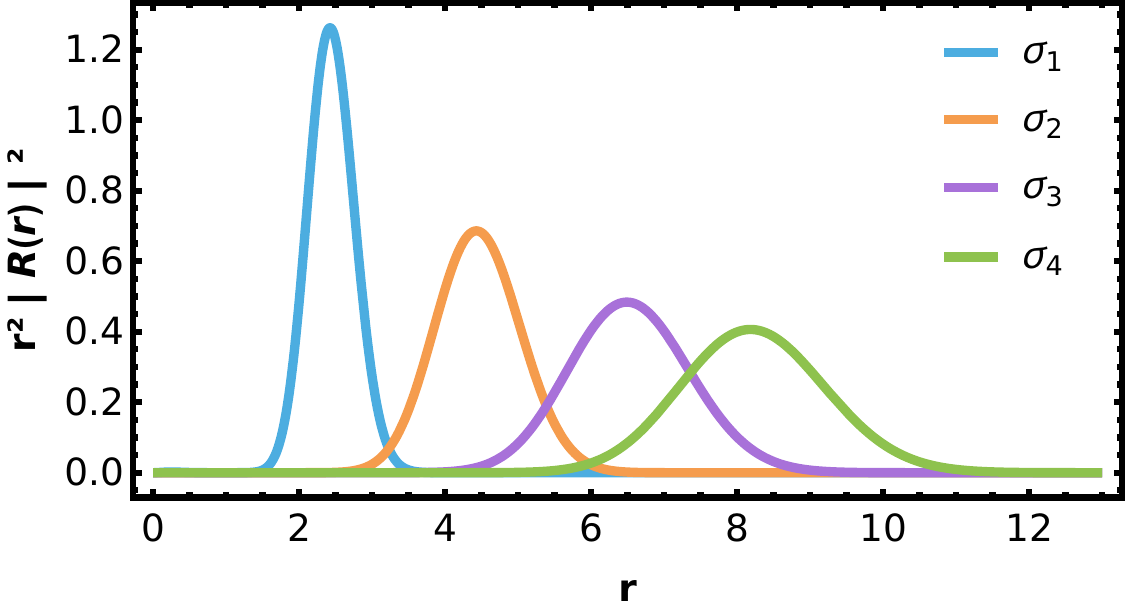}
  \caption{$M = -1$, $\mu_0 = 1$.}
  \label{fig:sfig-b}
\end{subfigure}
\begin{subfigure}{.5\textwidth}
  \centering
  \includegraphics[width=0.9\linewidth]{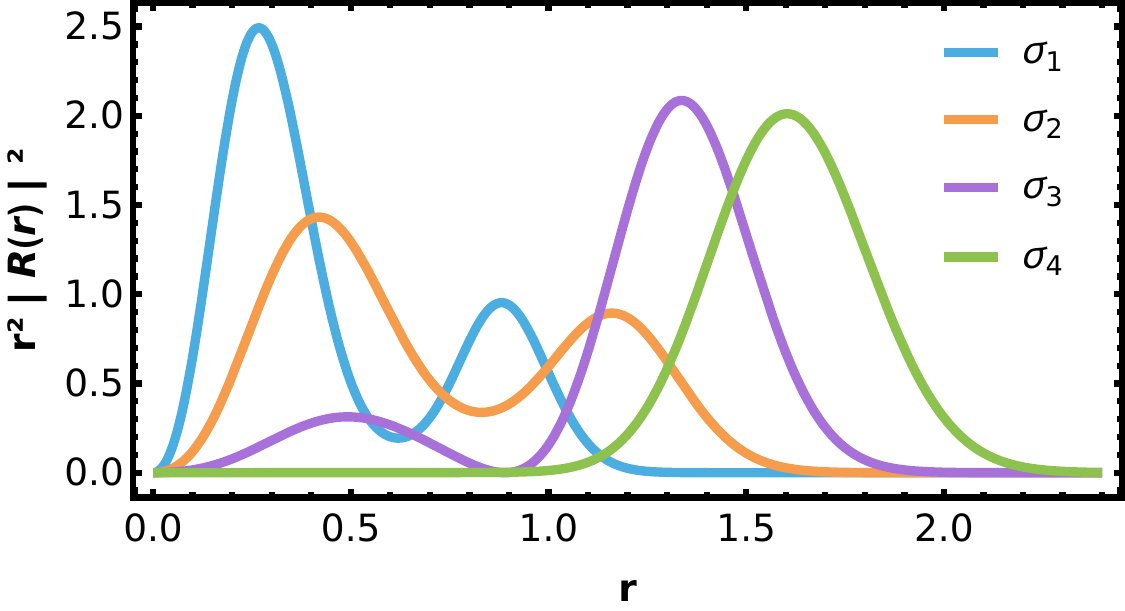}
  \caption{$M = -1$, $\mu_0 = 10$.}
  \label{fig:sfig-c}
\end{subfigure}
\begin{subfigure}{.5\textwidth}
  \centering
  \includegraphics[width=0.9\linewidth]{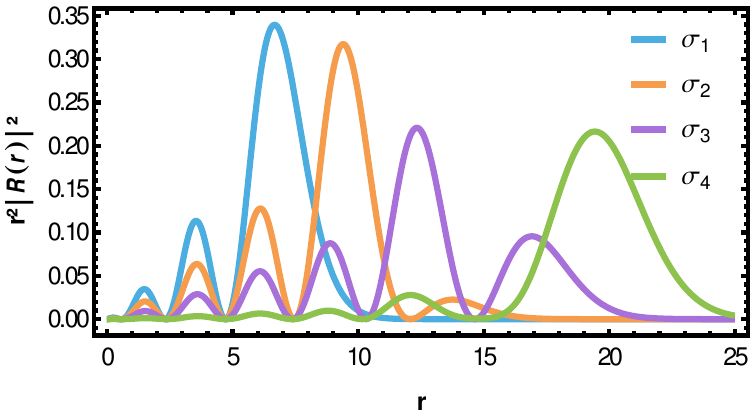}
  \caption{$M = 1$, $\mu_0 = 0.06$.}
  \label{fig:sfig-d}
\end{subfigure}
\begin{subfigure}{.5\textwidth}
  \centering
  \includegraphics[width=0.9\linewidth]{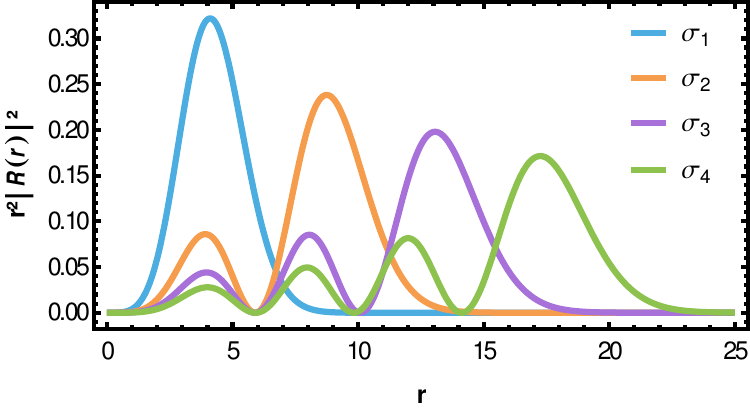}
  \caption{$M = 1$, $\mu_0 = 0.06$.}
  \label{fig:sfig-e}
\end{subfigure}
\begin{subfigure}{.5\textwidth}
  \centering
  \includegraphics[width=0.9\linewidth]{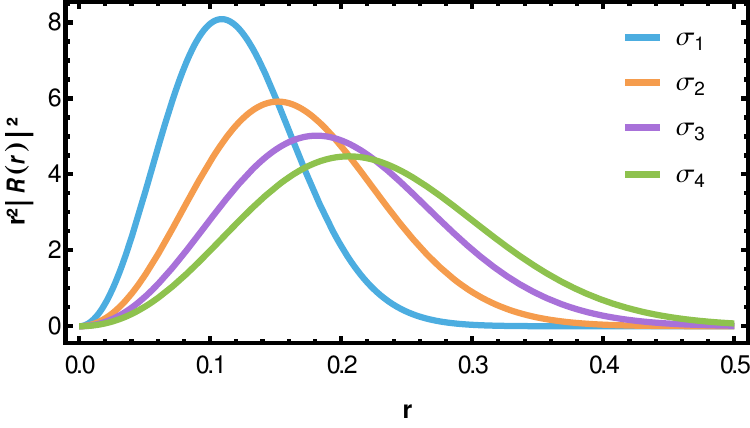}
  \caption{$M = -20$, $\mu_0 = 18$.}
  \label{fig:sfig-f}
\end{subfigure}
%
\caption{Probability dentities for the scalar field in the background of a static, chargeless black string.}
\label{fig:fig}
\end{figure}

Consider the line element \cite{griff} of a black string in a 3+1 dimensions
\be \label{line-element}
ds^2 = -Q dt^2 + \frac{1}{Q} dr^2 +r^2\left( d\rho^2 + \rho^2 d\phi^2\right), \ee
where $Q = \varepsilon - \frac{2M}{r} + \frac{q^2}{r^2} - \frac{\Lambda}{3}r^2$, where $M \neq 0$ is a real mass, $q$ is an electric charge and $\Lambda$ is the cosmological constant. If $\varepsilon = 0$, positive $M$ means the $r$-coordinate is time-like and the $t$-coordinate is space-like. On the other hand, their roles are inverted if $M$ is negative. Let us focus on the case where $\varepsilon = q = \Lambda = 0$. This case has been identified \cite{deep-interior} as the ``deep interior'' of the Schwarzschild black hole. The Klein-Gordon equation for a particle with mass $\mu_0$ is
\be \label{KGE}
\frac{1}{\sqrt{-g}} \; \partial_{\alpha}\left( g^{\alpha\beta} \sqrt{-g}\; \partial_\beta \right) \Psi - \mu_0^2\Psi =0, \ee
where $\sqrt{-\det g} = \sqrt{-g} = \rho r^2$. We substitute the contravariant components of the metric to obtain
\be\label{KG-antes}
-\frac{\sigma^2r^3}{4M^2} - \frac{1}{R}\frac{d}{dr}\left(r\frac{dR}{dr}\right) + \frac{1}{2M\rho S}\frac{d}{d\rho}\left(\rho\frac{dS}{d\rho}\right) - \frac{\nu^2}{2M\rho^2} - \frac{\mu_0^2r^2}{2M} = 0, 
\ee
where we introduced the ansatz,
\be\label{ansatz}
\Psi(t,r,\rho,\phi) = \exp\left( i\sigma t + i\nu\phi\right) R(r) S(\rho), \ee
in the above wave function $\sigma$ is the complex frequency, $\nu$ is the azimuthal quantum number and $R$ and $S$ are functions to be determined. We proceed with the standard approach by introducing a separation constant $\lambda$. The ``angular-"like equation reads,
\be\label{angular}
\frac{d^2S}{d\rho^2} + \frac{1}{\rho}\frac{dS}{d\rho} + \left( 2M\lambda  - \frac{\nu^2}{\rho^2}\right)S = 0 ,\ee
which is promptly recognized as the Bessel equation. The solutions are the well-known combination of ordinary Bessel functions and first kind Hankel function,
\be\label{solucao-angular}
S_\nu(\rho) = c_\nu J_\nu\left( \sqrt{2M\lambda}\rho\right) + d_\nu H_\nu^{(1)}\left( \sqrt{2M\lambda}\rho\right), \ee
where $\nu$ is an integer number. These cylindrical functions can be interpreted as representations for standing and propagating waves, respectively. The ``radial-"like equation is more interesting and also more complicated,
\be\label{radial}
\frac{d^2R}{dr^2} + \frac{1}{r}\frac{dR}{dr} + \left( \frac{\lambda}{r} + \frac{\mu_0^2 r}{2M} + \frac{\sigma^2r^2}{4M^2} \right)R = 0. \ee
One can transform this equation into the normal form, that is, without the first derivative, via $R(r) = g(r)/\sqrt{r}$ and, introducing $r=\kappa x$, with $\kappa^4 = -4M^2/\sigma^2$ the equation reads,
\be\label{radial-normal-canonica}
\frac{d^2g}{dx^2} + \left( \frac{\lambda \kappa}{x} + \frac{\mu_0^2\kappa^3 x}{2M} + \frac{1}{4x^2} -x^2 \right)g = 0. \ee
By means of the transformation $R(r) = r^A e^{Br}e^{Cr^2}f(r)$ we can also express equation \eqref{radial} in the canonical form,

\begin{equation}\label{eq-radial}
    \frac{d^2 f}{d r^2}+ \left(\frac{1}{r} +\frac{i \mu_0^2 }{\sigma }+\frac{i \sigma  r}{M}\right)\frac{d f}{d r} +\frac{1}{r}\left[\left(\frac{i \sigma
   }{M} - \frac{\mu_0^4}{4\sigma^2} \right)r + \lambda + \frac{i  \mu_0^2}{2\sigma}\right]f(r) = 0,
\end{equation}
where $A = 0$, $B_\pm = -\mu_0^2/(8 C_\pm M)$ and $C_\pm = \pm i \sigma/(4 M)$ and we choose $C = C_+$. Perusal of references \cite{ronveaux,slavyanov,olver} reveals that the differential equations \eqref{radial-normal-canonica} and \eqref{eq-radial} are the Heun's biconfluent equation. See equation (1.3.1) of part $D$, written by Maroni. The Heun's biconfluent equation was studied recently by Kamath \cite{pramana} is an interesting problem involving quantum coupled oscillators in the presence of a classical gravitational field, and by Karwowski and Witek \cite{harmonium} to model a three-body system known as  harmonium. It is possible to construct a local solution that is analytic at the origin by writing,
\begin{equation}\label{serie-heunb}
    f(r) = \text{HeunB}\left(q, \alpha, \gamma, \delta,\varepsilon , r\right) = \sum_{n = 0}^\infty a_n r^{n+s}.
\end{equation}
To complete our solution, we must follow the steps outlined by Kristensson \cite{kristensson} in order to calculate the allowed eigenvalues $\sigma$ that makes \eqref{serie-heunb} convergent for all $r < \infty$. Incidentally, upon doing so, we find that the eigenvalues lie in the complex numbers. This may indicate the presence of quasinormal modes for the generic case where $\varepsilon$, $q$ and $\Lambda$ may be different from zero, with the appropriate boundary conditions. The regular radial-like solution becomes

\begin{equation}
    R(r) = \exp{\left(\frac{i\mu_0^2}{2\sigma}r + {\frac{i \sigma}{4} r^2}\right)} \text{HeunB}\left(-\lambda -\frac{i  \mu_0^2}{2\sigma}, \frac{i \sigma
   }{M} - \frac{\mu_0^4}{4\sigma^2}, 1, \frac{i \mu_0^2}{\sigma },\frac{i \sigma
   }{M},r\right) \ .
\end{equation}
In Fig. \ref{fig:fig} we plot the non-normalized radial probability densities for the scalar particle with $\mu_0$ mass propagating in the black string background. In the (b) and (d) panels, we show the probability densities when $\lambda >0$, while all others are calculated setting $\lambda < 0$. Observe that even with unconventional values of the masses $M$ and $\mu_0$ the wavefunction is still convergent \footnote{The physical significance of the solutions are still a topic in dispute as noticed by Griffiths and Podolský \cite{griff}.}. In Table \ref{tab-sigma} we show the first four eigenvalues for the scalar particle in the black string background.
\begin{table}[H]
\centering
\begin{tabular}{ccccccc}
\multicolumn{7}{c}{Energy eigenvalues for the black string problem} \\ \hline
 &  &            &  &  & $\lambda = k_{0,1}$ & $\lambda = -k_{0,1}$ \\
 &  & $\sigma_1$ &  &  & $0.341773 i$        & $0.349022 i$         \\
 &  & $\sigma_2$ &  &  & $0.182980 i$        & $0.189057i$          \\
 &  & $\sigma_3$ &  &  & $0.127718 i$        & $0.133473 i$         \\
 &  & $\sigma_4$ &  &  & $0.098619i$         & $0.104211 i$         \\ \cline{2-7} 
\end{tabular}
\caption{The eigenvalues are calculated in units of $\hbar = c = G = 1$. The parameters are $M = 1$, $\mu_0 = 0.06$ and $k_{\ell,n}$ is the n-th zero of the Bessel function $J_\ell(x)$.}
\label{tab-sigma}
\end{table}

\section{Analogue black string}\label{secao3}
We can use the previous results to investigate a very interesting mapping between two distinct systems: the one we studied in the \ref{secao2} and an one-dimensional system studied by Kamath \cite{pramana}, namely, the two coupled oscillators interacting harmonically plus a classical gravitational potential. The \se{} for this problem reads, 
\be\label{schrodinger-kamath}
\left[ \frac{d^2}{dr^2} - 2\mu lr^2+ 4a\mu l r +\frac{2G\mu m^2}{r} +2\mu\left( E_n - 4a^2l\right) \right]\psi(r) = 0,
\ee
where $m$ is the oscillating particles mass, $l=3k/4$, $\mu$ is the system's reduced mass, $a$ is the system's equilibrium position, $G$ is Newton's constant, $r$ is the distance, $k$ the spring coupling constant, and $E_n$ is the energy eigenvalue. Introducing an external potential $V_{ext}(r)$ we obtain,
\be\label{schrodinger-kamath-pot-ext}
\left[ \frac{d^2}{dr^2} - 2\mu lr^2+ 4a\mu l r +\frac{2G\mu m^2}{r} +2\mu\left( E_n - V_{ext}(r)- 4a^2l\right) \right]\psi(r) = 0.
\ee
Let us call it {\it laboratory equation}, noticing that comparing it with the canonical biconfluent Heun equation, it is written in the normal form.

Now turn to our equation \eqref{radial-normal-canonica} and let us call it {\it target equation}. Both laboratory and target equations, i.e.,  \eqref{schrodinger-kamath-pot-ext} and \eqref{radial-normal-canonica} can be mapped to each other if the external potential could be adjusted to,
\be\label{Vext}
V_{ext}(r) = -\frac{1}{8\mu r^2} - 4a^2 l + E_n, \ee
namely, to an energy dependent constant potential plus an attractive inverse square potential. The change of variable $r = \xi x$, where $\xi = (2\mu l)^{-1/4}$, makes it straightforward to relate the laboratory parameters $l$ and $V_0$ to the target parameters $\lambda, \sigma, \mu_0,$ and $M$, since $\xi$ and $\kappa$ are inversely proportional. These relations are listed in table \ref{tabela-taquion}. So, adjusting the laboratory parameters to,
\be\label{parametros-lab}
m = \sqrt[3]{\frac{\lambda}{G}},\quad l = - \frac{4M^2}{\sigma^2} \sqrt[3]{\frac{G}{\lambda}}, \quad a =- \frac{\mu_0^2\sigma^2}{16 M^3},  \ee
the laboratory equation \eqref{schrodinger-kamath-pot-ext} can be transformed into \eqref{radial-normal-canonica}. The laboratory wavefunction obeys the same differential equation the target radial wavefunction satisfies. Since $m$ is a mass, $a$ is a length and $k$ is a positive constant the mapping can be achieved when $\sigma$ is a pure imaginary complex frequency.



\begin{table}[H]
\begin{tabular}{|c|c|}
\hline
Coupled oscillators plus $V_{Newton}+ V_{ext}$ (Lab)                                & Particle in the Tachyon background (Target)              \\ \hline
$\psi(r)$                                        & $ \sqrt{r} R(r)$           \\ \hline
$\xi$ & $1/\kappa$ \\ \hline

$Gm^3 $              & $\lambda$      \\ \hline
$ 2aml$ & $ \mu_0^2/(2M)$ \\ \hline
\end{tabular}

\caption{Mapping between physical quantities which enter radial equations \eqref{schrodinger-kamath-pot-ext} and \eqref{radial-normal-canonica}. In the first/second row we list physical quantities for the laboratory/target system, respectively. In the first line we write the eigenfunctions, and in the next three the parameters that multiply $r^{-1}, r,$ and $r^2$, respectively. }
\label{tabela-taquion}
\end{table}

\section{Conclusions}\label{secao4}
We studied the Klein-Gordon equation for a massive particle interacting with the gravitational field of a kind of topological black hole called black string, holding a mass $M$. The equation is separable, and we managed to solve it exactly. We presented the wavefunction as a product of complex exponentials $\exp(i\sigma t+i\nu\phi)$ times cylindrical functions --- Bessel and first kind Hankel functions of $\rho$ --- times a biconfluent Heun's function, which depends on $r$. The probability densities are shown in figure \ref{fig:fig}, and they illustrate that there are no normal modes for the particle in this background. We contribute to the analogue gravity initiative suggesting that a damped quantum harmonic oscillator as an analogue for the particle in the chargeless, non-rotating black string background in a static universe. 

As we suggest in this work, the emergence of imaginary eigenvalues may indicate the presence of quasinormal modes and other effects when we consider charge --- whether electric, magnetic, or dyonic --- and the cosmological constant.


\appendix
\section{The Biconfluent Heun Differential Equation}

 The biconfluent Heun differential equation (BHE) can be achieved as follows \cite{slavyanov}. Starting from the confluent Heun equation (CHE)
 \begin{equation}
     y'' + \left(\frac{\gamma}{z} + \frac{\delta}{z-1} + \varepsilon\right)y' + \left[\frac{\alpha z - q}{z(z-1)}\right]y(z) = 0,
 \end{equation}
 we shift the singular point $z = 1$ to $z = \frac{1}{b}$ through $z = b z'$ and $y(z(z')) = f(z')$. Then, imposing $\alpha = \alpha'/b^2$, $q = q'/b$, and $\varepsilon = -\varepsilon'/b^2$, together with $\delta = -\delta'/b - \varepsilon'/b^2$, we take the limit $b\rightarrow\infty$ and drop the primes, acquiring 
\begin{equation}\label{equacao-heunb}
    f'' + \left(\frac{\gamma}{z} +\delta + \varepsilon z\right)f' + \left(\frac{\alpha z - q}{z}\right)f(z) = 0,
\end{equation}
 which we refer as the canonical form of the BHE. This form is somewhat in agreement with \cite{ronveaux}, although our BHE is more general, and it has a $+$ instead of a $-$ sign in the second term, contrasting with the NIST Handbook \cite{olver}. This notation is also in agreement with \cite{pramana} and it can make use of the Mathematica standard notation for its solution, $\text{HeunB}(q,\alpha,\gamma,\delta,\varepsilon,r)$.

 To solve \eqref{equacao-heunb}, we introduce the ansatz
\begin{equation}\label{ansatz-heunb-generalizada}
    f(z) = \sum_{n=0}^\infty a_n z^{n+s},
\end{equation}
from which, upon substitution in \eqref{equacao-heunb}, results in the three-term recurrence relation
\begin{gather}
        a_0 s(s + \gamma - 1) = 0, \label{eq para os valores de s}\\
        a_0 (s\delta - q) + a_1 (s+1)(s+\gamma) = 0, \label{eq para a1 e a0}\\
        a_{n+1} (n+1)(n+\gamma) + a_n (n\delta - q) + a_{n-1}(\gamma + (n-1)\varepsilon) = 0 \label{relacao de 3 termos}
\end{gather}
From equation \eqref{eq para os valores de s}, we have the values of s, namely, $s = {0,1-\gamma}$. From \eqref{eq para a1 e a0}, results
\begin{equation*}
    a_1 = \frac{q - \delta s}{(s + 1)(s + \gamma)}a_0,
\end{equation*}
which means that we must impose the restriction $\gamma \neq 0, -1, -2, ...$. To proceed further, we massage equation \eqref{relacao de 3 termos} into the shape
\begin{equation}\label{relacao de recorrencia kristensson}
    a_{n+1} = \underbrace{\frac{(q - n\delta)}{(n+1)(n+\gamma)}}_{A_n}a_n  \overbrace{- \frac{(\gamma + (n-1)\varepsilon)}{(n+1)(n+\gamma)}}^{B_n}a_{n-1} = A_n a_n + B_n a_{n-1} .
\end{equation}
We are now prepared to tackle this three-term recurrence relation with the methodology laid down by Kristensson \cite{kristensson}, in which we rewrite
\begin{equation*}
    a_1 = \frac{q - \delta s}{(s + 1)(s + \gamma)}a_0 = A_0 a_0,
\end{equation*}
and, as Kristensson concludes, we are allowed to consider

\begin{equation}
    a_n = D_n a_0,
\end{equation}
with $D_{n+1}$ being the determinant

\begin{equation}
D_{n+1} = 
\begin{vmatrix}
A_0 & -1 & 0 & 0 & \cdots & 0 & 0 \\
B_1 & A_1 & -1 & 0 & \cdots & 0 & 0 \\
0 & B_2 & A_2 & -1 & \cdots & 0 & 0 \\
\vdots & \vdots & \vdots & \ddots & \ddots & \vdots & \vdots \\
0 & 0 & 0 & \cdots & B_{n-1} & A_{n-1} & -1 \\
0 & 0 & 0 & \cdots & 0 & B_{n} & A_{n}
\end{vmatrix}
\end{equation}
The power series is convergent so long as the limit 

\begin{equation}
    \lim_{n\rightarrow\infty} D_n = 0
\end{equation}
is satisfied. This equation can be understood as an eigenvalue equation, that gives us a condition on $q$ to unsure the convergence of \eqref{ansatz-heunb-generalizada}. The convergence can be determined by observing the limit
\begin{equation*}
    \lambda = \lim_{n\rightarrow\infty} \frac{a_{n+1}}{a_n} = \lim_{n \rightarrow\infty} A_n +B_n \frac{a_{n-1}}{a_n} = A + \frac{B}{\lambda},
\end{equation*}
that exists according to the Poincaré--Perron theorem, where $A = \lim_{n \rightarrow\infty} A_n$ and $B = \lim_{n \rightarrow\infty} B_n$. We see that

\begin{align*}
    A_n &\approx -\frac{\delta}{n}+\frac{\delta(
   \gamma + 1)+q}{n^2}+O\left(\frac{1}{n^3}\right) \\
   B_n &\approx -\frac{\epsilon}{n}+\frac{\epsilon \gamma+2
   \epsilon-\gamma}{n^2}+O\left(\frac{1}{n^3}\right).
\end{align*}
Consequently, we find $\lambda_1 = \lambda_2 = 0$, meaning the radius of convergence is infinite.
Another possibility lies in the polynomial solutions of BHE. To achieve these solutions, we must turn back to \eqref{relacao de recorrencia kristensson} and set $a_n = B_n = 0$. To do so, we notice that if $\gamma/\varepsilon = 1 - N$, where $N \in \mathbb{N}$ and $a_n = D_n = 0$, we have $a_{n+1} = a_n = 0$, thus truncating the series \eqref{ansatz-heunb-generalizada}.
 \section*{Acknowledgments}
	
The authors gratefully acknowledge CNPq (grant numbers 309052/2023-8 and 140471/2022-7) for partial financial support.





\end{document}